\documentclass{aastex}
\usepackage{emulateapj5,natbib}

\newcommand{\hr}{\mbox{$^h$}}
\renewcommand{\min}{\mbox{$^m$}}

\renewcommand{\deg}{\mbox{$^{\circ}$}}

\newcommand{\nvw}{\mbox{\ion{N}{5} $\lambda$1240}}

\newcommand{\civw}{\mbox{\ion{C}{4} $\lambda$1549}}
\newcommand{\heiiw}{\mbox{\ion{He}{2} $\lambda$1640}}
\newcommand{\ciiiw}{\mbox{\ion{C}{3}] $\lambda$1909}}
\newcommand{\neivw}{\mbox{[\ion{Ne}{4}] $\lambda$2424}}

\newcommand{\niiaw}{\mbox{[\ion{N}{2}] $\lambda$6548}}
\newcommand{\niibw}{\mbox{[\ion{N}{2}] $\lambda$6583}}

\newcommand{\oiiibw}{\mbox{[\ion{O}{3}] $\lambda$5007}}

\newcommand{\lya}{\mbox{Ly$\alpha$}}

\newcommand{\siiv}{\mbox{\ion{Si}{4}}}

\newcommand{\nii}{\mbox{[\ion{N}{2}]}}
\newcommand{\hal}{\mbox{H$\alpha$}}
\newcommand{\sii}{\mbox{[\ion{S}{2}]}}

\newcommand{\wrst}{\mbox{$W_{\lambda, \mbox{\tiny rest}}$}}
\newcommand{\haln}{\mbox{H$\alpha$(n)}}
\newcommand{\halb}{\mbox{H$\alpha$(b)}}
\newcommand{\simgtr}{\; \raisebox{-.2ex}{$\stackrel{>}{\mbox{\tiny $\sim$}}$} \;}
\newcommand{\simlt}{\; \raisebox{-.2ex}{$\stackrel{<}{\mbox{\tiny $\sim$}}$} \;}
\newcommand{\sigsky}{\mbox{$\sigma_{\mbox{\tiny sky}}$}}

\slugcomment{\it Accepted for publication in the March 2003 Astronomical Journal.}

\lefthead{Dawson et al.}
\righthead{A High--Redshift, Hard X--ray Emitting Spiral}

\begin{document}

\title{Optical and Near--Infrared Spectroscopy of a High--Redshift,
Hard X--ray Emitting Spiral Galaxy\altaffilmark{1}}

\author{
Steve Dawson\altaffilmark{2},
Nate McCrady\altaffilmark{2},
Daniel Stern\altaffilmark{3},
Megan E. Eckart\altaffilmark{4},
Hyron Spinrad\altaffilmark{2},
Michael C. Liu\altaffilmark{5},
and James R. Graham\altaffilmark{2}
}

\altaffiltext{1}{
Based on observations made at the W.M. Keck Observatory, which is operated
as a scientific partnership among the California Institute of Technology,
the University of California and the National Aeronautics and Space
Administration.  The Observatory was made possible by the generous
financial support of the W.M. Keck Foundation.}

\altaffiltext{2}{
Department of Astronomy, University of California at Berkeley, Mail Code
3411, Berkeley, CA 94720 USA; sdawson@astro.berkeley.edu,
nate@astro.berkeley.edu, spinrad@astro.berkeley.edu,
jrg@astro.berkeley.edu.}

\altaffiltext{3}{
Jet Propulsion Laboratory, California Institute of Technology, Mail Stop
169--327, Pasadena, CA 91109 USA; stern@zwolfkinder.jpl.nasa.gov.}

\altaffiltext{4}{
Division of Physics, Mathematics, and Astronomy, California Institute of
Technology, Mail Stop 220--47, Pasadena, CA 91125 USA;
eckart@srl.caltech.edu.}

\altaffiltext{5}{
Currently Beatrice Watson Parrent Fellow, the Institute for Astronomy,
University of Hawai'i, 2680 Woodlawn Drive, Honolulu, HI 96822 USA;
mliu@IfA.Hawaii.Edu.}


\begin{abstract}
We present optical and near--infrared Keck spectroscopy of CXOHDFN
J123635.6$+$621424 (hereafter HDFX28), a hard X--ray source at a redshift
of $z = 2.011$ in the flanking fields of the Hubble Deep Field--North
(HDF--N). HDFX28 is a red source (${\cal R} - K_s = 4.74$) with extended
steep--spectrum ($\alpha^{\mbox{\tiny 8.4 GHz}}_{\mbox{\tiny 1.4 GHz}} >
0.87$) microjansky radio emission and significant emission (441 $\mu$Jy)
at 15 $\mu$m. Accordingly, initial investigations prompted the
interpretation that HDFX28 is powered by star formation.  Subsequent {\it
Chandra} imaging, however, revealed hard ($\Gamma = 0.30$) X--ray emission
indicative of absorbed AGN activity, implying that HDFX28 is an obscured,
Type II AGN.  The optical and near--infrared spectra presented herein
corroborate this result;  the near--infrared emission lines cannot be
powered by star formation alone, and the optical emission lines indicate a
buried AGN. HDFX28 is identified with a face--on, moderately late--type
spiral galaxy. Multi--wavelength morphological studies of the HDF--N have
heretofore revealed no galaxies with any kind of recognizable spiral
structure beyond $z > 2$. We present a quantitative analysis of the
morphology of HDFX28, and we find the measures of central concentration
and asymmetry to be indeed consistent with those expected for a rare
high--redshift spiral galaxy.
\end{abstract}



\section{Introduction}
\label{introduction}

The origin of the X--ray background (XRB) remains an enduring puzzle for
X--ray astronomy.  Great progress has been made during the last four
years: {\it ROSAT} surveys successfully resolved $\sim$80\% of the soft
XRB (0.5--2 keV) into discrete sources \citep[e.g.][]{hasinger98} and
current work with the {\it Chandra X--ray Observatory} (hereafter, {\it
Chandra})  has successfully resolved a similar fraction of the hard XRB
\citep[2--8 keV;][]{brandt01, giacconi01, giacconi02, hornschemeier01,
rosati02}.  Nonetheless, a coherent understanding of the physical and
evolutionary properties of the sources which comprise the XRB is only just
now emerging. Although much of this population appears to be the nuclei of
otherwise normal bright galaxies ($I < 23.5$) or typical active galactic
nuclei (AGNs), a significant fraction of the discrete sources is optically
faint ($I > 23.5$), and therefore not easily identified
\citep[e.g.][]{alexander01, barger02}.  {\it Type II quasars}, for
instance, are thought to be AGNs viewed edge--on through an obscuring
torus \citep{antonucci93} and are deemed an essential component of the
XRB--producing population \citep{moran01}. However, few well--studied
examples of such systems are known at high redshift, and owing in part to
their lack of relativistic brightening, they are not easily identified in
shallow, large--area surveys \citep{norman02, stern02q}.

Type II quasars represent just one of several diverse classes of objects
emerging in follow--up work to deep {\it Chandra} fields
\citep[e.g.][]{hornschemeier01, schreier01, stern02}. On the
extra--galactic side, we find X-ray--loud composite galaxies typified by
starburst or early--type optical spectra which bear no signature of their
buried AGN \citep[e.g.][]{moran96, levenson01, stern02}. Additionally, we
find X--ray sources whose optical counterparts belong to the class of
faint, extremely red objects (EROs), the nature of which has remained
uncertain owing to the difficulty in spectroscopic follow--up
\citep[e.g.][]{alexander02, elston88, elston89, hu94, graham96, liu00,
hornschemeier01, stern02}. On the Galactic side, we find late--type dwarfs
emitting soft X--rays originating in chromospheric activity
\citep[e.g.][]{hornschemeier01}, and very low mass binary systems emitting
hard X--rays driven by accretion \citep[e.g.][]{stern02}.  Amidst the
emergence of this menagerie of objects, optical and near--infrared
spectroscopic follow--up has become increasingly vital not only to
identifying the source population of the XRB, but also to elucidating the
physics of X--ray sources in general, and to delineating their evolution
with redshift.

\begin{figure}
\centering
\epsscale{1.0}
\plotone{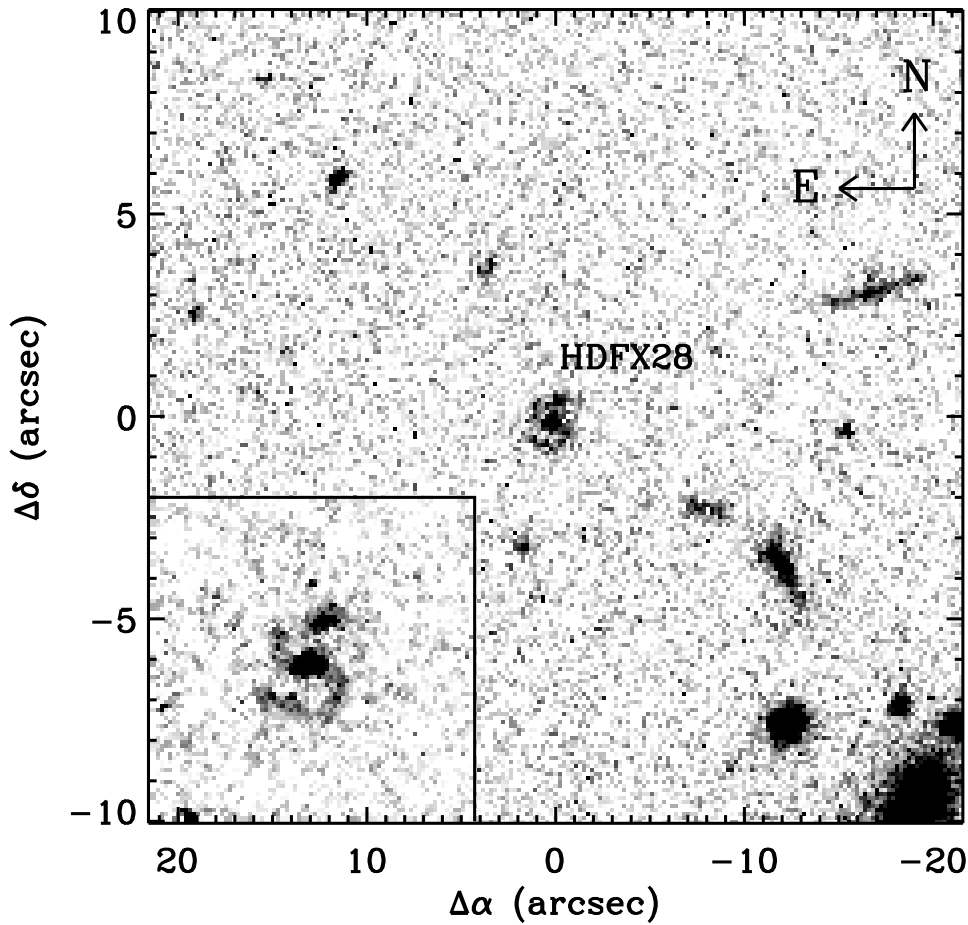}
\caption{Central portion of the single--orbit {\it HST} $I_{814}$ Hubble
Deep Field inner west flanking field \citep{williams96}, centered on HDFX28 at $\alpha =
12\hr36\min35\farcs6$, $\delta = +62\deg14\arcmin24\arcsec$ (J2000). The panel
measures 20\arcsec\ square and the orientation is indicated.
The inset shows a 4\arcsec\ square subsection of a
two--orbit preliminary image from the first epoch of the Great Observatory
Origins Deep Survey \citep[GOODS;][]{dickinson02}
{\it HST} Treasury Program (L. Moustakas 2002, private
communication).  The image was taken on UT 2002 Nov 21 
with the Advanced Camera for Surveys \citep[ACS;][]{pavlovsky01} and is the sum of 0.5--orbit
$V_{606}$ integration, a 0.5--orbit $I_{775}$ integration, and a 1.0--orbit $z_{850}$
integration.  With repeat {\it HST}
visits through June 2003, the GOODS program will increase the ACS
integration on this field five--fold.
}
\label{flank}
\end{figure}

One critical facet of this endeavor is simply to distinguish between
objects powered by mass accretion onto supermassive black holes (quasars
and other AGNs) and those powered by nuclear fusion in stars (normal and
starburst galaxies).  To this end, we present optical and near--infrared
spectra of CXOHDFN J123635.6$+$621424 (hereafter HDFX28), a hard X--ray
source identified with a face--on spiral galaxy at redshift $z = 2.011$
(Figure~\ref{flank}). HDFX28 is fortuitously located in the Hubble Deep
Field--North inner west (HDF--N IW) flanking field, and was therefore
subject to a vast array of follow--up imaging. As such, HDFX28 was
initially identified as an extended microjansky radio source with a
comparatively steep spectral index ($S_{\nu} \propto \nu^{-\alpha}$;
$\alpha^{\mbox{\tiny 8.4 GHz}}_{\mbox{\tiny 1.4 GHz}} > 0.87$).  Together
with its detection by the {\em Infrared Space Observatory} Camera (ISOCAM)
and its pronounced optical spatial extent ($\sim 1\farcs6$), the radio
data for HDFX28 prompted an initial interpretation as a galaxy powered by
star formation \citep[e.g.][]{richards00}. However, as we discuss below,
the detection of HDFX28 as a hard X--ray source in the deep {\it Chandra}
survey of the HDF--N \citep{hornschemeier01, brandt01}, corroborated by
the spectroscopy presented herein, demonstrates that this galaxy in fact
harbors an obscured, Type II AGN.

In addition to confirming its AGN status, the spectroscopy of HDFX28
indicates a surprisingly high redshift for an object with identifiable
spiral structure.  \citet{dickinson00} summarizes the results of
morphological studies of the HDF--N by reporting a total lack of even
plausible candidates for spiral galaxies at $z>2$.  Prompted by this lack
of precedent for high--redshift spirals, we present a quantitative study
of central concentration and asymmetry in HDFX28 based on the scheme
devised by \citet{abraham96} for the analysis of large CCD imaging
surveys. With the application of a modest morphological $k$--correction,
we find HDFX28 to have morphological parameters consistent with those
derived from catalogs of both artificially redshifted nearby spirals, as
well as catalogs of {\it HST} imaging of spirals out to $z \sim 1$
\citep{abraham96}.

In short, HDFX28 is intriguing both for its membership in the emerging
class of X--ray--selected Type II AGN, and for possessing a morphology
which is unprecedented at its redshift. We describe the optical and
near--infrared spectroscopy of HDFX28 in section \S \ref{observations},
and we present the results of the spectroscopy and the classification of
the source as an obscured, Type II AGN in \S \ref{as_an_agn}.  We report
on our quantitative analysis of its morphology in \S \ref{as_a_spiral},
and we summarize our results in \S \ref{conclusion}. Throughout this paper
we adopt the currently favored $\Lambda$--cosmology of
$\Omega_{\mbox{\tiny M}} = 0.35$ and $\Omega_\Lambda = 0.65$, with $H_0 =
65$ km s$^{-1}$ Mpc$^{-1}$ \citep[e.g.][]{riess01}.  At $z=2.011$, such a
universe is 3.22 Gyr old, the lookback time is 76.9\% of the total age of
the Universe, and an angular size of 1\farcs0 corresponds to 8.66 kpc.

\bigskip
\bigskip
\bigskip

\section{Spectroscopic Observations}
\label{observations}

\subsection{Optical Spectroscopy}
\label{optspectroscopy}

We obtained the optical spectrum of HDFX28 on UT 2001 February 23 as part
of an observing campaign of photometrically--selected high--redshift
candidates in the HDF--N and its environs \citep[][]{dawson01}.  The data
were taken with the Low Resolution Imaging Spectrometer
\citep[LRIS;][]{oke95} at the Cassegrain focus on the 10m Keck I
telescope, after the advent of the LRIS--B spectrograph channel
\citep{mccarthy98}.  The red--sensitive LRIS--R camera uses a Tek 2048$^2$
CCD detector with a pixel scale of 0\farcs212 pixel$^{-1}$; the
blue--sensitive LRIS--B camera is nearly identical.  The data were taken
with slitmasks designed to obtain spectra for $\sim 15$ targets
simultaneously through 1\farcs5 wide slits. For this observation, we used
the 400 lines mm$^{-1}$ grating blazed at 8500 \AA\ (1.86 \AA\, pix$^{-1}$
dispersion) in the red channel, and a 300 lines mm$^{-1}$ grism blazed at
5000 \AA\ (2.64 \AA\, pix$^{-1}$ dispersion) in the blue channel.  To
split the red and blue channels, we used a dichroic with a cutoff at 6800
\AA. With this setup, the combined spectrograph channels afforded a
spectral coverage of roughly 3200 \AA\ to 1 $\mu$m, covering the entire
optical window.  The total exposure time of 2.75 hours was broken into
three exposures of 1500 seconds and three exposures of 1800 seconds;  we
performed $\sim 3\arcsec$ spatial offsets between exposures to facilitate
the removal of fringing at long wavelengths.  The airmass never exceeded
1.75 during the observations.

We used the IRAF\footnote{IRAF is distributed by the National Optical
Astronomy Observatories, which are operated by the Association of
Universities for Research in Astronomy, Inc., under cooperative agreement
with the National Science Foundation.} package \citep{tody93} to process
the slitmask data, following, where possible, standard slit spectroscopy
procedures.  One deviation from standard spectroscopic data reduction was
that the blue--channel spectrum was not divided by a flatfield exposure.
This omission is a consequence of the fact that the existing internal
halogen lamp produces no light shortward of 3800 \AA, and also appears to
contain prominent UV emission lines.  Our experience with attempts at
flatfielding LRIS--B in a variety of spectroscopic set--ups indicates that
the pixel--to--pixel variations corrected by flatfielding are typically $<
4$\%, and that they have little systematic variation across the CCD; as
such, we expect that our flux--calibrated spectra are little affected by
this treatment. Remaining aspects of treating the slitmask data were
facilitated by a home--grown software package, BOGUS\footnote{BOGUS is
available online at
http://zwolfkinder.jpl.nasa.gov/$\sim$stern/homepage/bogus.html.}, created
by D. Stern, A.J.  Bunker, and S.A. Stanford.

\begin{figure}
\centering
\epsscale{1.0}
\plotone{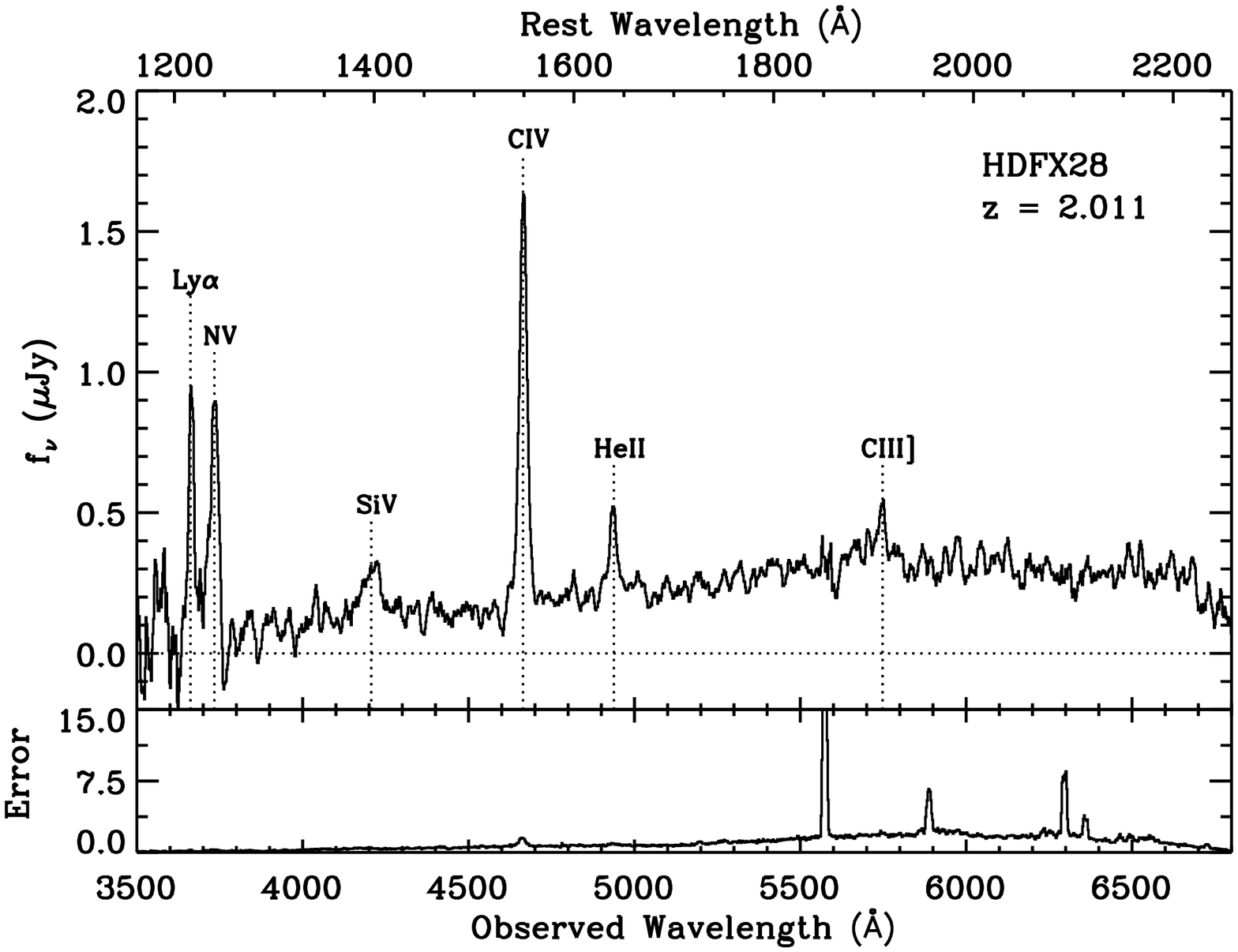}
\caption{(Top) Blue channel optical spectrum of HDFX28 obtained with
LRIS--B on the Keck I telescope.  The spectrum was extracted using the
optimal extraction algorithm described in \citet{horne86}, and was
smoothed with a boxcar filter of length equal to one resolution element
($\Delta \lambda \sim 14$ \AA\ at $\lambda = 5000$ \AA, based on Gaussian
fits to night--sky emission lines). The total integration time was 2.75
hours. (Bottom) The statistical uncertainty per pixel over the same
wavelength range and in the same flux units as the object spectrum.
}
\label{optspecB}
\end{figure}

\begin{figure}
\centering
\epsscale{1.0}
\plotone{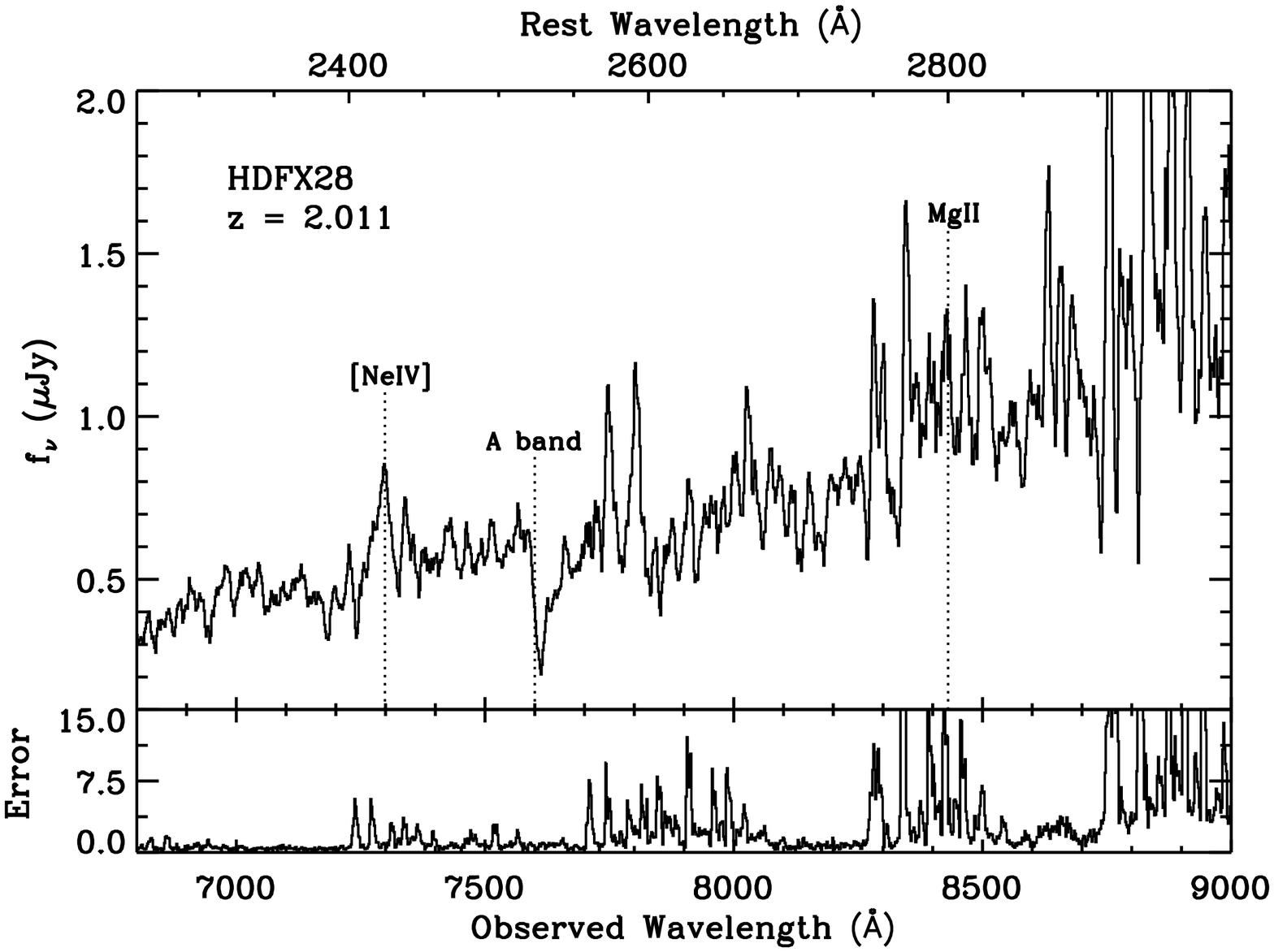}
\caption{(Top) Red channel optical spectrum of HDFX28 obtained with
LRIS--R on the Keck I telescope.  The spectrum was extracted using the
optimal extraction algorithm described in \citet{horne86}, and was
smoothed with a boxcar filter of length equal to one resolution element
($\Delta \lambda \sim 11$ \AA\ at $\lambda = 8000$ \AA, based on Gaussian
fits to night--sky emission lines). The total integration time was 2.75
hours. Note that the unlabeled, narrow spectral features longward
of 7800 \AA\ are artifacts due to imperfect subtraction of telluric OH and
O$_2$ night--sky emission lines, and that the absorption feature at $7600$
\AA\ is the telluric A--band. (Bottom) The statistical uncertainty per
pixel over the same wavelength range and in the same flux units as the
object spectrum.
}
\label{optspecR}
\end{figure}

We extracted the blue--channel and red--channel spectra using the optimal
extraction algorithm described in \citet{horne86}.  Wavelength
calibrations were performed in the standard fashion using Hg, Ne, Ar, and
Kr arc lamps; we employed telluric sky lines to adjust the wavelength
zero--point.  We performed flux calibrations with longslit observations of
standard stars from \citet{massey90} taken with the instrument in the same
configuration as the multislit observation. However, it should be noted
that owing to the constraints of observing with a slitmask, the data were
taken at a position angle of 163.1\deg, not at the parallactic angle. The
final extracted blue--channel optical spectrum is shown in
Figure~\ref{optspecB}; the final extracted red--channel optical spectrum
is shown in Figure~\ref{optspecR}.

\subsection{Near--Infrared Spectroscopy}
\label{irspectroscopy}

We obtained the near--infrared spectrum of HDFX28 with the 10m Keck II
telescope on UT 2001 April 13, using the facility near--infrared
spectrometer NIRSPEC \citep{mclean98}.  We employed a 0\farcs57 $\times$
42\arcsec\ slit to achieve low resolution ($R \sim 1300$) spectra in the
wavelength range 1.75--2.17 $\mu$m.  We obtained four 600 second
integrations, with $\sim 5\arcsec$ spatial offsets between exposure. The
data were dark subtracted, flat--fielded and corrected for cosmic rays and
bad pixels in the standard fashion.  We sky--subtracted by pairwise
subtraction of successive nods along the slit.  The curved spectral order
was then rectified onto a slit--position/wavelength grid based on a
wavelength solution from arc lamp emission lines.  As data from the second
slit position had significantly lower signal-to-noise, likely the result
of temporary seeing degradation or misalignment of the slit, they were
rejected.  The total integration time for the near--infrared spectrum is
thus 1800 seconds.

\begin{figure}
\centering
\epsscale{1.0}
\plotone{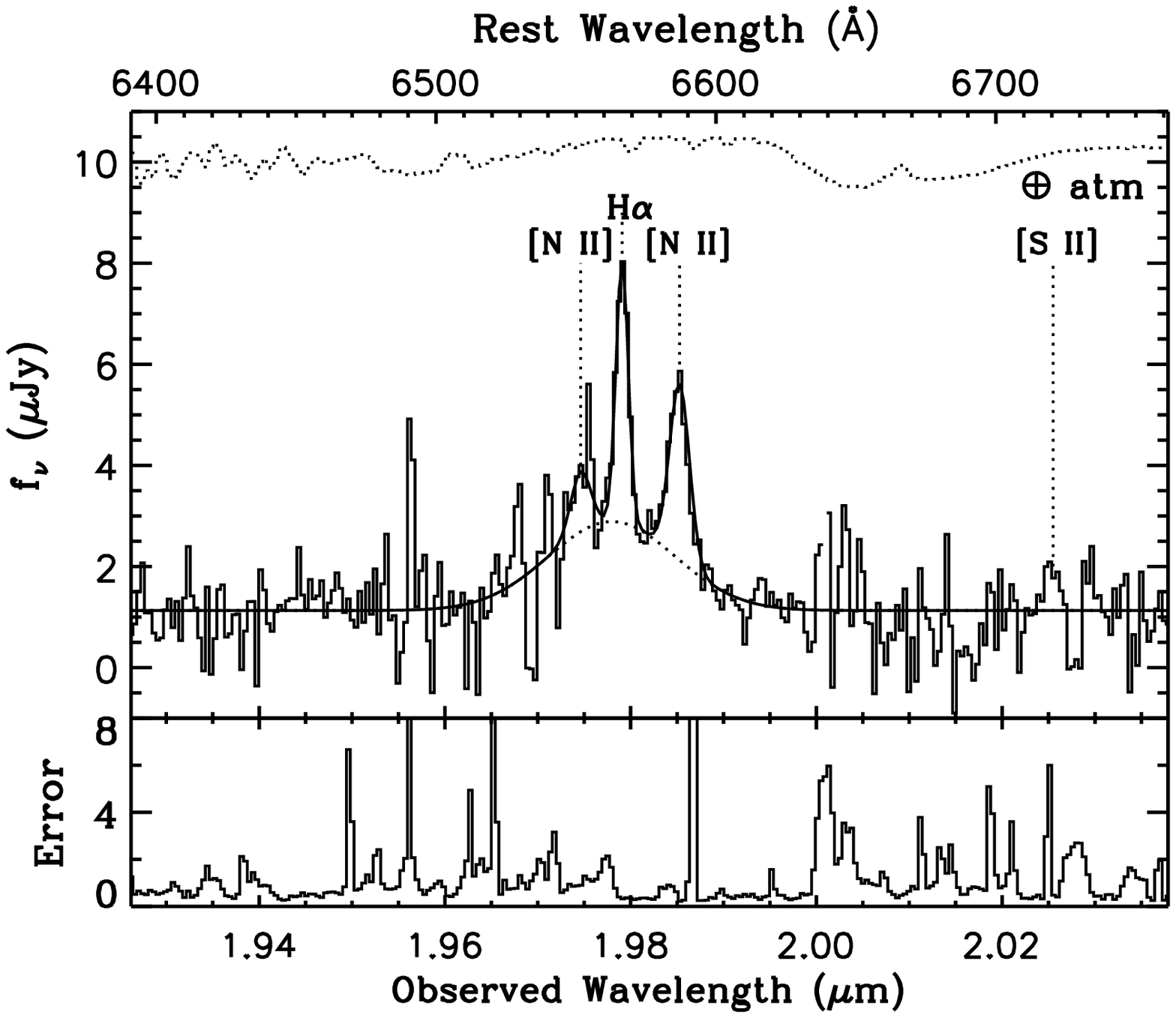}
\caption{(Top) Near--infrared spectrum of HDFX28 obtained with NIRSPEC on
the Keck II telescope.  Four Gaussians were fit to the the emission
complex. The total fit (solid curve) consists of \niiaw, \niibw, and a
narrow \hal\ component, superposed on a broad \hal\ component (dotted
curve).  The unresolved \sii\ 6716 \AA\ / 6731 \AA\ doublet is barely
discernible at $\sim 2.025$ $\mu$m; it was not included in the fit.  The
dotted line at the top of the plot shows terrestrial atmospheric
absorption, arbitrarily scaled.  (Bottom) The calculated error in each
wavelength bin in the same units as the object spectrum. The dominant
source of error is sky subtraction; the peaks are due to bright sky
emission lines which were not well subtracted.
}
\label{irspec}
\end{figure}

The galaxy spectrum was extracted using a Gaussian weighting function
which was matched to the wavelength--integrated profile.  To correct for
atmospheric absorption, we divided the galaxy spectrum by the spectrum of
an A0V calibration star, HD 99966.  Both the galaxy and the star were
observed at an airmass of $\sim 1.3$.  The resulting spectrum is shown in
Figure~\ref{irspec}.

\section{HDFX28 as a Type II AGN}
\label{as_an_agn}

\subsection{Results from the Optical Spectrum}
\label{optresults}

The optical spectrum of HDFX28 shows resolved, moderate--width,
high--ionization emission lines typical of AGNs (Figures~\ref{optspecB}
and \ref{optspecR}).  Both the permitted and forbidden lines are
well--identified, allowing for unambiguous determination of the redshift.
To this end, and to ascertain the fluxes and widths of the emission lines,
we made a weighted, single Gaussian Levenberg--Markwardt fit to each
isolated line\footnote{In the case of the unresolved \siiv\ doublet, we
fit with two Gaussians with amplitudes constrained by the ratio of the
doublet Einstein $A$--values, 1.02:1; due to the large uncertainty in the
fit, however, this line was not used in any of the following analysis.},
resulting in a redshift of $z = 2.011$.  We note that this value deviates
somewhat from that presented in a recent spectroscopic catalog of {\it
Chandra} sources in the HDF--N \citep[source 142, $z=2.00$;][]{barger02}.
The emission line parameters are cataloged in Table~\ref{lines}.

The emission line widths in the optical spectrum present a solid case for
the classification of HDFX28 in the overall taxonomy of AGN. The canonical
definition of a Seyfert 1 galaxy involves a spectrum with broad permitted
lines, typically $\simgtr 5000$ km s$^{-1}$ FWHM, and comparatively narrow
forbidden lines, typically $\sim 500$ km s$^{-1}$ FWHM
\citep[e.g.][]{osterbrock89}.  The definition of a Seyfert 2, by contrast,
involves a spectrum showing permitted and forbidden lines of approximately
the same FWHM, typically $\sim 500$ km s$^{-1}$. On this account, the
rough agreement between the widths of the permitted and forbidden UV lines
in HDFX28 calls for classification as a Type II source.  Moreover, though
the permitted line widths of $\simgtr 1000$ km s$^{-1}$ slightly exceed
those expected for a prototypical Seyfert 2, they still fall far short of
permitted line widths observed in Type I sources, or in the broad line
regions of Type II sources seen in polarized light
\citep[e.g.][]{vernet01}. In particular, the width of the \heiiw\ line
compares favorably with the nine high--redshift radio galaxies (HzRGs)
presented in \citet{vernet01}; HzRGs are perhaps the best--studied class
of obscured, Type II AGN at the redshift of HDFX28
\citep[e.g.][]{mccarthy93, eales93, eales96, evans98}.  Furthermore, the
line widths of HDFX28 compare favorably with those reported for Type II
AGN elsewhere in the literature: e.g.\ $\sim 900$ km s$^{-1}$ for the
infrared--selected Type II quasar IRAS 09104$+$4109 \citep{kleinmann88};
$\sim 1000$ km s$^{-1}$ for the Type II quasar in the {\it Chandra} Deep
Field South, CDF--S 202 \citep{norman02}; and $\simgtr 1000$ km s$^{-1}$
for the Type II quasar in the deep {\it Chandra} Lynx field, CXO52
\citep{stern02q}.

\begin{figure}
\centering
\epsscale{1.0}
\plotone{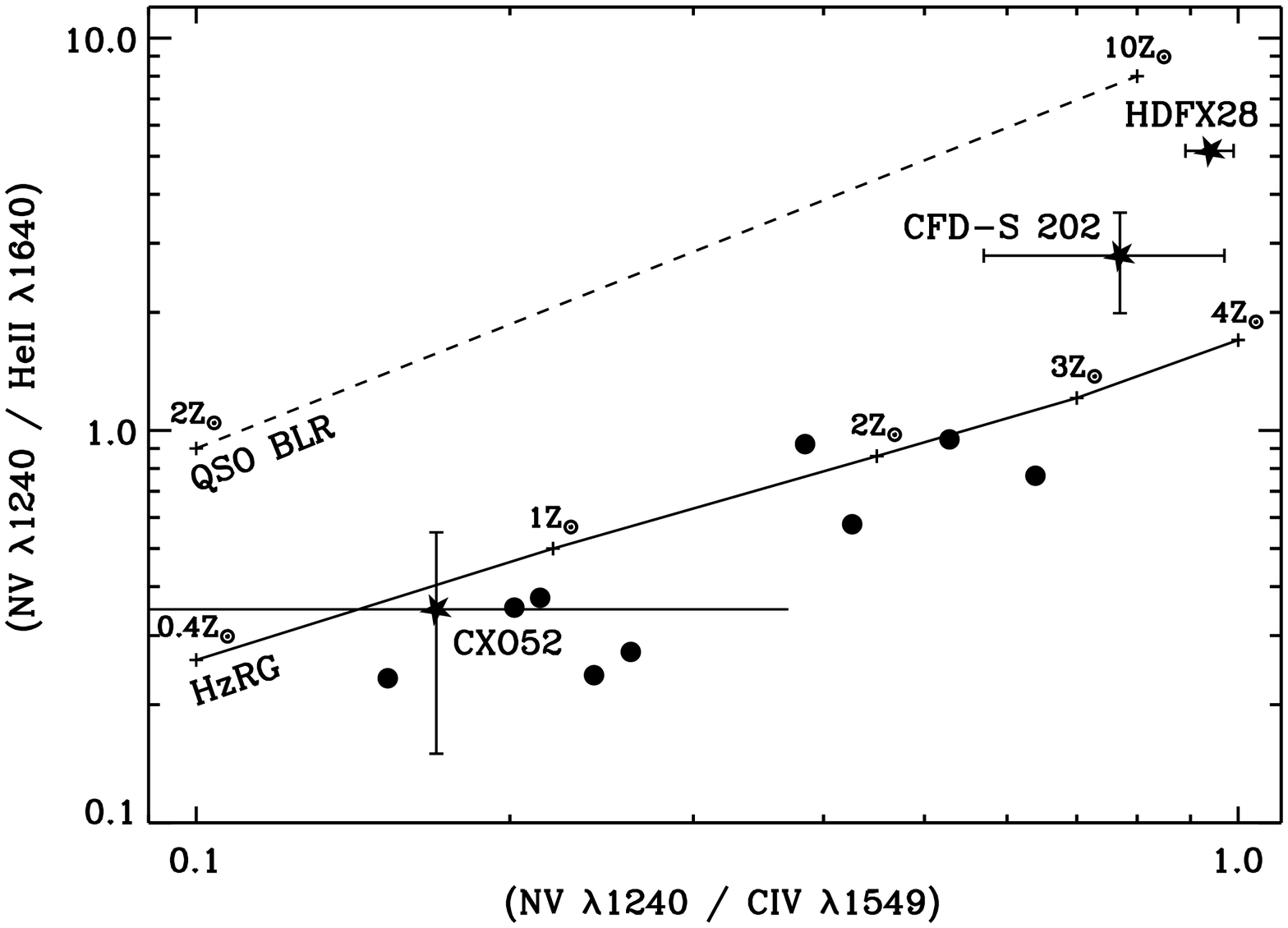}
\caption{The \ion{N}{5} $\lambda$1240 / \ion{C}{4} $\lambda$1549 vs.
\ion{N}{5} $\lambda$1240 / \ion{He}{2} $\lambda$1640 plane as it appears
in \citet{vernet01} and \citet{norman02}, showing HDFX28 along with the
Type II quasars CDF--S 202 \citep{norman02} and CXO52 \citep{stern02q}.
The labeled stars indicate the Type II sources. The circles indicate nine
high--redshift radio galaxies (HzRGs) presented by \citet{vernet01}; the
HzRGs are the only class of Type II AGN which have been studied
extensively at the redshift of HDFX28. The dashed line represents the
locus of a QSO broad--line region (BLR) chemical evolution model with
metallicities ranging from 2--10 times solar \citep{hamann93}. The solid
line indicates the locus of the best fit power--law photoionization models
for HzRGs with metallicities ranging from 0.4--4 times solar
\citep{vernet01}. HDFX28 occupies a position intermediate between the two
models, and is evidently of high metallicity.
}
\label{ratioplane}
\end{figure}

The emission line flux ratios for HDFX28, however, may somewhat weaken the
case for classification as a straight Type II source.  We tabulate the
flux ratios for HDFX28 along with those of two other high--redshift Type
II AGNs in Table~\ref{ratios}, and we plot the sources in the \nvw\ /
\heiiw\ vs.  \nvw\ / \civw\ plane in Figure~\ref{ratioplane}. Both these
ratios are comparatively strong in HDFX28, placing it intermediate between
models for the narrow emission lines of HzRGs and models for QSO
broad--line regions (BLRs), though closer to the QSO BLRs. Moreover,
though the location of HDFX28 in Figure~\ref{ratioplane} compares
favorably to that of CDF--S 202, HDFX28 is far stronger in both flux
ratios than CXO52. Of course, the \nvw\ emission in CXO52 was noted as
exceptionally weak; \citet{stern02q} report that its \nvw\ / \civw\ ratio
is approximately half of what is seen in composite HzRG spectra
\citep[e.g.][]{mccarthy93,stern99hzrg}. Nonetheless, the comparative
strength of \nvw\ in HDFX28 may point to a classification intermediate
between Type I and Type II AGN.

The \civw\ / \heiiw\ ratio for HDFX28 indicates a similar conclusion.
Typical values of \civw\ / \heiiw\ for unobscured, Type I objects are
$\sim 10$ in composite UV spectra of Seyfert 1 galaxies \citep{heckman95},
and 7--50 in composite quasar spectra \citep{boyle90, francis91,
vandenberk01}. Typical values for obscured, Type II objects are $\sim 1$
in composite UV spectra of Seyfert 2 galaxies \citep{heckman95}, and $\sim
1.5$ in composite spectra of HzRGs \citep{mccarthy93, stern99hzrg}.  On
this account, HDFX28 is again intermediate between the Type I and Type II
sources, though it is worth noting that the other two high--redshift
objects in Table~\ref{ratios} somewhat echo this trend, and both are
nevertheless classified as Type II AGN. Still, particularly in
anticipation of the weak, broad \hal\ emission described below (\S
\ref{irresults}), we conclude that the optical spectrum of HDFX28 favors
classification somewhere on the continuum between Type I and Type II
sources, rather than as prototypically Type II.

On a separate note, the \lya\ line of HDFX28 is exceptionally weak both in
equivalent width and in relative flux. For comparison, \citet{mccarthy93}
offers a mean rest--frame equivalent width for HzRGs at $z > 1.5$ of
$\wrst ( \lya ) = 295 \pm 188$ \AA, and \citet{stern99hzrg} give a mean
rest--frame equivalent width for 17 HzRGs spanning $0.3 < z < 3.6$ of
$\wrst ( \lya ) = 75$ \AA.  Whereas the high--ionization state emission
lines of HDFX28 have equivalent widths similar to those reported elsewhere
\citep[e.g. $\sim 10$--$10^2$ \AA\ for CXO52;][]{stern02q}, we find for
HDFX28 a meager $\wrst ( \lya ) = 35 \pm 3$ \AA. The weakness of \lya\ in
HDFX28 in relative flux is dramatic both observationally and
theoretically.  Relative to \civw\ and \nvw, \lya\ emission in both CDF--S
202 and CXO52 exceeds that in HDFX28 by factors ranging from $\sim 2$ to
10. Furthermore, \citet{ferland85} predict an unreddened \lya\ / \hal\
ratio of 16 from a model spectrum of a classical Seyfert 2 galaxy.
Anticipating our near--infrared results (below), we report a \lya\ / \hal\
ratio for HDFX28 of just $2.4 \pm 0.1$.

With $E(B-V) = 0.00$ towards the HDF--N \citep{williams96}, Galactic
extinction is unviable as a culprit for the diminished \lya\ flux in
HDFX28.  Rather, as has been deduced from weak \lya\ in several HzRGs
\citep[e.g.][]{eales93,dey95}, it is likely that dust in HDFX28 is
preferentially extinguishing \lya\ photons. Since \lya\ is a resonant
line, \lya\ photons are multiply scattered by neutral hydrogen as they
traverse the system, resulting in a long path length for dust absorption.
Hence, it is possible for \lya\ emission to become substantially depressed
even if there is little dust in the system, so long as there is sufficient
neutral hydrogen.

\subsection{Results from the Near--Infrared Spectrum}
\label{irresults}

The near--infrared spectrum of HDFX28 shows weak continuum emission and a
resolved emission line complex near 1.98 $\mu$m identified as \hal\ plus
[\ion{N}{2}] $\lambda \lambda 6548, 6583$ \AA. As is common in the spectra
of Seyfert galaxies, the \hal\ emission consists of two components: a
strong narrow line superposed upon a weak, broad line
\citep[e.g.][]{osterbrock89}.  We therefore fit the emission complex with
four Gaussians subject to the following constraints: (1) the ratio of
amplitudes of the \nii\ doublet lines must be 2.96:1 as prescribed by the
ratio of their Einstein $A$--values, (2) the \nii\ lines must have the
same redshift as the narrow \hal\ component, and (3) the two \nii\ lines
must have identical FWHMs.  Again, we made a Levenberg--Markwardt fit to
the spectrum, weighting according to the uncertainty of each pixel. The
resulting fit is overlaid on the spectrum in Figure~\ref{irspec}, and the
corresponding emission line parameters are cataloged along with the
results from the optical spectroscopy in Table \ref{lines}.

Like the optical spectrum discussed above, the near--infrared spectrum
indicates a somewhat mixed result for the AGN classification of HDFX28.
The widths of the \nii\ doublet lines and of the narrow component of the
\hal\ emission are comparable to the narrow lines of a classic Seyfert 2
\citep[e.g.][]{osterbrock89}, and they compare favorably with both the
forbidden and permitted optical line widths reported for CXO52 in
\citet{stern02q}.  However, these results must be mitigated by the
presence of weak, broad \hal\ emission.  We find a FWHM of $2500 \pm 250$
km~s$^{-1}$ for this broad \hal\ component, and a ratio of
broad--to--narrow emission of $\halb / \haln = 3.3 \pm 0.3$.  For
comparison, the broad component to the \hal\ emission in HDFX28 is far
weaker than that of the HzRG MRC 2025--218 \citep{larkin00}, with its
\halb\ FWHM of $9300 \pm 900$ km~s$^{-1}$ and broad--to--narrow flux ratio
of $\halb / \haln = 7 \pm 2$. \citet{stern02q} point out that, though the
spectrum of MRC 2025--218 over the range from \lya\ to \oiiibw\ is very
similar to classic obscured, Type II AGN observed by other groups
\citep[e.g.][]{eales93, eales96, evans98}, the presence of broad \hal\ may
indicate that MRC 2025--218 is actually the high--luminosity analog to a
Seyfert 1.8, rather than a Seyfert 2. In keeping with the results of the
optical spectroscopy, the same claim may therefore be made for HDFX28.

One further diagnostic offered by the near--infrared spectrum of HDFX28 is
the ratio of its \niibw\ flux to its \haln\ flux.  \citet{veilleux87} and
\citet{osterbrock89} present a classification scheme employing these and
other optical features to discriminate the narrow lines of AGNs from those
of starburst galaxies.  The physical distinction exploited in this case is
the differing strengths of low--ionization lines such as \niibw\ in each
class of source. In the narrow line region of an AGN, these
low--ionization lines arise preferentially in an extended zone of partly
photoionized hydrogen which results from an ionizing spectrum containing a
large fraction of high--energy photons.  These photons are absent in the
spectrum of OB stars; hence, the strength of the low--ionization lines is
diminished in starburst galaxies \citep[e.g.][]{baldwin81, veilleux87,
osterbrock89}.  Typically, starbursts and \ion{H}{2} region--like galaxies
occupy $-2 \simlt \log (\niibw / \haln) \simlt -0.3$, while AGN occupy
$-0.3 \simlt \log (\niibw / \haln) \simlt 0.8$ With reference to
Table~\ref{ratios}, we find HDFX28 to fall definitively within the regime
of AGNs.

This result is corroborrated by the rest--frame equivalent width of the
near--infrared emission complex.  Based on an average continuum level of
$1.13 \pm 0.06$ $\mu$Jy, the total equivalent width of the (\hal\ + \nii)
feature is $\wrst = 150 \pm 40$~\AA.  By contrast, surveys of normal (i.e.
non--AGN) galaxies find an average width in the range of just 20--30 \AA\
\citep[e.g.][]{kennicutt83}, and surveys of starburst galaxies find only
$\sim 40$ \AA\ \citep[e.g.][]{ravindranath01}.

\subsection{Results from the Multi--Wavelength Photometry}
\label{photometry}

The aim of the {\it Hubble Space Telescope} ({\it HST}) observations of
the HDF--N was to image an otherwise undistinguished field as deeply as
reasonably possible \citep{williams96}.  Indeed, the HDF--N represents the
deepest optical images ever taken, providing detections and photometry of
stars and field galaxies to $V \sim 30$ with 0\farcs1 resolution, and
reaching source densities of $\sim 10^6$ deg$^{-2}$ \citep[for a review,
see][]{ferguson00}. One caveat is that the {\it HST} images of the HDF--N
are rather small, covering only $\sim 5$ arcmin$^2$.  Hence, to facilitate
ground--based follow--up observations, the deep imaging program was
augmented with short, 1--2 orbit images of eight fields immediately
adjacent to the primary field.  These flanking field observations were
made exclusively with the WFPC2 $I_{814}$ filter \citep[][Table
2]{williams96}.

HDFX28 is located 1.6\arcmin\ west and 1.4\arcmin\ north of the pointing
center of the primary HDF--N in the inner west flanking field. Its optical
counterpart gives the impression of a moderately late--type face--on
spiral galaxy measuring roughly 1\farcs6 (14 kpc) in diameter
(Figure~\ref{flank}). For their program of associating {\em Infrared Space
Observatory} ({\it ISO})  detections with optical sources in and adjacent
to the HDF--N, \citet{mann97} constructed an $I_{814}$ catalog of the
flanking fields; they give $I_{814} = 23.46$ for HDFX28.

Owing to its location in the HDF--N flanking fields, HDFX28 has been
inadvertantly subject to a panoply of follow--up imaging (see
Table~\ref{photo}). The galaxy was first reported as a weak radio source
(8.15 $\mu$Jy at 8.5 GHz; 87.8 $\mu$Jy at 1.4 GHz) in the sensitive radio
surveys of \citet{richards98} and \citet{richards00}, respectively. These
results yield a comparatively steep radio spectral index ($S_{\nu} \propto
\nu^{-\alpha}$; $\alpha^{\mbox{\tiny 8.4 GHz}}_{\mbox{\tiny 1.4 GHz}} >
0.87$), with radio emission extending across 2\farcs8.  In general,
microjansky radio emission from disk galaxies can result from either star
formation (e.g.\ from free--free emission originating in \ion{H}{2}
regions) or from AGN activity connected with a central engine.
\citet{richards00} argued that (1) in the case of a central AGN powering a
weak ($P < 10^{25}$ W Hz$^{-1}$)  radio source, the bulk of the radio
emission is confined to the nuclear region and is therefore characterized
by sub--arcsecond angular scales, and (2) such small scales result in a
high opacity to synchrotron self--absorption, yielding flat or inverted
spectral indices typically in the range $-0.5 < \alpha < 0.5$.  HDFX28 is
indeed a weak radio source, with $L^{\mbox{\tiny rest}}_{\mbox{\tiny 1.4
GHz}} \sim 2 \times 10^{24}$ W Hz$^{-1}$.  Operating under the assumption
that HDFX28 has a redshift in the range $0.2 < z < 1$ (as was inferred
from the spatial extent of HDFX28 in the flanking field image, and was
consistent with the bulk of the microjansky radio sources in deep VLA
surveys), \citet{richards00} would have estimated even less radio power.
Hence, the origin of the radio emission in HDFX28 was taken to be extended
star--forming regions.

This conclusion was ostensibly borne out by the ISOCAM detection of HDFX28
\citep{aussel99}.  If HDFX28 were a starburst galaxy at moderate--to--low
redshift, then the ISOCAM 15 $\mu$m filter (LW3) would sample rest
wavelengths from roughly 6 $\mu$m to 12 $\mu$m.  The mid--infrared
emission could therefore be plausibly attributed to the unidentified
infrared bands (UIB) and to the hot, 200 K dust which typically dominates
the spectral energy distribution of starbursts over those wavelengths
\citep{aussel99}.  Together, the radio and mid--infrared data therefore
appeared to paint a coherent picture of HDFX28 as a source with star
formation in its disk as its underlying emission mechanism.  This
conclusion was consistent with the general observation from deep VLA
surveys that the bulk of the radio population at the microjansky level
consists of starforming disks, with fewer than 20\% of the radio sources
associated with early--type galaxies or quasars \citep{fomalont96,
richards98}.

Subsequent ground--based optical and near--infrared imaging found HDFX28
to be comparatively red, falling just short of the conventional definition
of EROs (e.g.\ ${\cal R} - K_s > 5.0$; \citealt{hornschemeier01}, and $I -
K_s > 4$; \citealt{stern02}). \citet{hogg00} give ${\cal R} - K_s = 4.74$
for HDFX28, and \citet{barger00} add $I - K_s = 3.89$. In contrast to the
conclusion drawn from the radio and IR imaging discussed above, and in
anticipation of the X--ray data discussed below, it is notable that HDFX28
corroborates the trend that X--ray sources at modest $R$ band magnitudes
tend to be redder than typical field galaxies, and that in the X--ray
population there appears to be an excess of sources with $4 < {\cal R} -
K_s \simlt 5$ \citep[see][Figure 7]{hornschemeier01}.  It is yet more
notable that \citet{hasinger99} reports that all X--ray counterparts with
$(R - K^{\prime}) > 4.5$ in the ROSAT Ultra Deep HRI Survey are either
members of high--redshift clusters or are obscured AGNs.

On that note, the 1 Ms {\it Chandra} survey of the HDF--N and its environs
yielded a soft X--ray flux of $0.28 \times 10^{-15}$ ergs cm$^{-2}$
s$^{-1}$ and a hard X--ray flux of $2.82 \times 10^{-15}$ ergs cm$^{-2}$
s$^{-1}$ for HDFX28 \citep{brandt01}.  Together with the optical imaging,
these results correspond to an X--ray to optical flux ratio\footnote{
\citet{hornschemeier01} use the Kron--Cousins $R$ filter transmission
function to derive the X--ray to optical flux ratio: $\log ( f_X / f_R ) =
\log f_X + 5.50 + R / 2.5$.} of $\log ( f_X / f_R ) = -0.65$ in the soft
band and $\log ( f_X / f_R ) = 0.35$ in the hard band.  As noted by
\citet{hornschemeier01} and \citet{stern02}, the majority of X--ray
sources in shallow surveys fall within $-1 < \log ( f_X / f_R ) < 1$, as
this range is typical of local AGN.  In particular, these values compare
very favorably to the Type II QSO CDF--S 202 \citep{norman02}, with its
soft band ratio of $\log ( f_X / f_R ) = -0.61$ and its hard band ratio of
$\log ( f_X / f_R ) = 0.29$.

The X--ray data indicate that HDFX28 is a comparatively hard source.
Following the nomenclature of \citet{stern02}, the hardness ratio for
HDFX28 is $HR = 0.24 \pm 0.10$, comparing favorably to $HR = 0.07 \pm
0.13$ for the Type II quasar CXO52. \citet{brandt01} report an X--ray band
ratio for HDFX28 of $1.66^{+0.37}_{-0.30}$ and a corresponding estimate of
the photon index\footnote{The photon index $\Gamma$ is derived from a
power law model for the X--ray spectrum:  $N = AE^{-\Gamma}$, where $N$ is
the number of photons s$^{-1}$ cm$^{-2}$ keV$^{-1}$ and $A$ is a
normalization constant \citep[e.g.][]{hornschemeier01}.} of $\Gamma =
0.30$. These results show HDFX28 to be distinctly hard for its soft band
count rate compared to the total samples in both the {\it Chandra} survey
of the HDF--N \citep{brandt01} and the {\it Chandra} survey of the Lynx
field \citep{stern02}. Moreover, this photon index is quite unlike the
steep $\Gamma \sim 1.7$ -- $2.0$ indices typical of unobscured AGNs
\citep[e.g.][]{nandra94}.  By assuming that HDFX28 has an intrinsic power
law spectrum with $\Gamma = 1.8$ such that the observed band ratio is due
to obscuration at the source, and by adopting a Galactic absorption column
density in the direction of the HDF--N of $N_{\mbox{\tiny H}} = 1.7 \times
10^{20}$ cm$^{-2}$ \citep{williams96}, we estimate the hydrogen column
density at the source to be $N_{\mbox{\tiny H}} \sim 1.5 \times 10^{23}$
cm$^{-2}$.  This value places HDFX28 very near to the median
$N_{\mbox{\tiny H}}$ of the sample of 73 nearby Seyfert II galaxies
compiled by \citet{bassani99}, and it implies an unobscured full band
rest--frame luminosity of $1.1 \times 10^{44}$ erg s$^{-1}$, which is well
within the quasar regime. In short, each of these results point to
significant soft X--ray absorption by intervening material within HDFX28.
Hence, as first suggested by \citet{hornschemeier01} and as confirmed by
the optical and near--infrared spectroscopy presented herein, the X--ray
data show HDFX28 to be an obscured, Type II AGN.

\section{HDFX28 as a High--Redshift Spiral Galaxy}
\label{as_a_spiral}

It is surprising to find identifiable spiral structure at the early time
indicated by the redshift of HDFX28.  Careful, multi--wavelength
morphological studies of the HDF--N reveal no galaxies with any kind of
recognizable spiral structure at $z > 2$ \citep{dickinson00}.  To wit, the
redshift distribution of a sample of 52 late--type spiral and irregular
galaxies complete to $K < 20.47$ shows a dramatic cut--off at $z \sim
1.4$, with only two galaxies in the sample exceeding this limit
\citep{rodighiero00}.  Similarly, a combined photometric redshift /
morphological data set complete to $I < 26.0$ shows a sharp drop in the
spiral galaxy distribution at $z > 1.5$ \citep{driver98}.  Specifically,
in the $22 < I_{\mbox{\tiny AB}} < 23$ magnitude bin, there are no spiral
galaxies beyond $z > 1.5$; in the $23 < I_{\mbox{\tiny AB}} < 24$
magnitude bin, there is only one.

\citet{abraham94,abraham96} cautioned that visual morphological
classifications of late--type galaxies fainter than $I = 21$ are somewhat
subjective, particularly for distant systems with small image sizes. This
difficulty is most pernicious for {\it very} late spirals (morphological
type $T > 7$), and for merging systems and peculiar galaxies. Nonetheless,
especially when combined with the lack of precedent for spiral galaxies at
the redshift of HDFX28, this caveat prompted us to bolster our
qualitative, visual classification with a quantitative, objective
classification.

To this end, we employed a morphological classification scheme devised by
\citet{abraham96} for analysis of the current generation of large CCD
imaging surveys, and modified by \citet{kuchinski01} for patchy, low
signal--to--noise data.  The classification scheme is a two--part system
which uses quantitative measurements of the galaxy central concentration
and asymmetry to distinguish three morphological bins: E/S0 galaxies,
spiral galaxies, and irregular or peculiar systems.  The concentration
index ($C$) is the ratio of the light emitted from a central region of the
galaxy (usually $R < 0.3 R_{\mbox{\tiny max}}$, where $R_{\mbox{\tiny
max}}$ is the radius of an elliptical aperture centered on the galaxy) to
the light emitted from the galaxy as a whole. The asymmetry index ($A$) is
a measure of the $180\deg$ rotational symmetry of the galaxy, measured by
rotating the galaxy image about the central pixel and subtracting the
rotated image from the original image. In essence, galaxies with high
degrees of central concentration and symmetry have regular, ordered
appearances, roughly corresponding to early to mid--Hubble types.
Galaxies with low central concentration and large asymmetry have irregular
or peculiar morphologies, corresponding to late to irregular Hubble types.

We calculated $C$ and $A$ for HDFX28 using the definitions given by
equations (1) and (3) in \citet{kuchinski01}.  It has been shown that
these indices are sensitive to the definition of the center of the galaxy
image and to the aperture in which the indices are measured \citep[][and
references therein]{kuchinski01}.  As such, we determined the central
pixel by first smoothing the galaxy image with a Gaussian kernel of
$\sigma = 1$ pixel and then taking the location of the maximum pixel as
the galaxy center.  We defined the aperture by setting a threshold at $1.0
\sigsky$ and then defining an ellipse based on the intensity--weighted
moments of the resulting image. Both these methods have their precedent in
a significant body of similar work \citep[e.g.][]{abraham96,teplitz98}; we
effected the aperture definition with the source extraction software
package SExtractor \citep{bertin96}.  For the concentration index, we
found $\log C = -0.43 \pm 0.03$; for the asymmetry index, we found $\log A
= -0.37 \pm 0.07$. As we discuss below, these objective results are indeed
consistent with our qualitative classification of HDFX28 as a spiral
galaxy.

We estimate the uncertainty in $C$ and $A$ by considering two independent
sources of error: the statistical error due to Poisson noise entering into
the calculations ($\sigma_{\mbox{\tiny P}}$), and the variance introduced
by calculating $C$ and $A$ in apertures extending to different limiting
surface brightnesses ($\sigma_{\mbox{\tiny S}}$).  For $C$, the
uncertainty due to noise was determined by propagating the Poisson noise
per pixel through the calculation in the standard fashion; we found
$\sigma_{\mbox{\tiny P,$C$}} = 0.021$.  The uncertainty inherent in using
apertures defined to different limiting surface brightnesses was
quantified by calculating $C$ for 11 apertures of decreasing size
determined by running SExtractor with detection thresholds spanning $0.5
\, \sigsky$ to $1.5 \, \sigsky$.  The standard deviation of these
measurements was $\sigma_{\mbox{\tiny S,$C$}} = 0.026$.  Adding these
uncertainties in quadrature yielded our total uncertainty estimate of
$\sigma_{\mbox{\tiny $C$}} = 0.03$.

To calculate $A$, the absolute value is taken of the difference between
the original image and the rotated image, resulting in sky noise which
systematically contributes only positive values.  As in \citet{abraham96},
we corrected for this effect by subtracting from $A$ the measured
asymmetry of many ($10^2$) blank patches of sky with apertures equal to
that enclosing the galaxy.  At the same time, we estimated the Poisson
error by measuring the distribution of the asymmetry indices of these
sky--only apertures. This process yielded $\sigma_{\mbox{\tiny P,$A$}} =
0.058$.  The uncertainty due to using apertures defined by different
limits was determined for $A$ exactly as it was determined for $C$ with
the result $\sigma_{\mbox{\tiny S,$A$}} = 0.037$.  Together, these
considerations yielded a total estimated error of $\sigma_{\mbox{\tiny
$A$}} = 0.07$.

To interpret our results in terms of Hubble types, we compare HDFX28 to
two large reference samples for which the concentration and asymmetry
indices have been calculated: (1) the \citet{frei96} catalog of nearby
galaxies artificially redshifted to $z = 0.3$, 0.5, and 0.7
\citep[][figure 2]{abraham96}, and (2) the catalog of galaxies imaged in
the {\it HST} Medium Deep Survey (MDS), which are expected to have a
redshift distribution spanning $0 < z < 1.0$ with a peak at $z = 0.6$
\citep[][figures 5 and 6]{abraham96}.  Our initial impression is that the
central concentration index of HDFX28 is very typical of spiral galaxies,
but that the asymmetry index straddles the border in $A$ between the
spirals (low $A$) and the peculiars (high $A$). However, as HDFX28 is at a
higher redshift than the most distant objects in either of these samples,
proper interpretation of its morphological indices requires that we first
consider the effects of cosmological distances on morphology.

Three effects complicate the issue of morphology for galaxies at
high--redshift:  bandshifting, surface brightness dimming, and the loss of
spatial resolution. Based on a careful study of {\it in situ} ultraviolet
and optical imaging of 32 local galaxies, \citet{kuchinski01} report that
bandshifting is the dominant effect.  As such, the general trend is for
$C$ to decrease and $A$ to increase as one proceeds to higher redshifts.
Among other effects, at shorter rest wavelengths apparent morphology
becomes dominated by localized star formation, thereby diminishing the
effect of an optical bulge (if any) on $C$ and increasing the patchiness
measured by $A$. As for HDFX28, we note that the effect of bandshifting on
$C$ is less pronounced in later spiral galaxies that lack dominant bulges
to begin with; the Sbc--Sd spirals in the \citet{kuchinski01} sample show
an average move of $\Delta C \sim 0.1$, where $\Delta C = C_{\mbox{\tiny
OPT}} - C_{\mbox{\tiny FUV}}$.  Hence, we would expect little change in
$C$ if we were able to observe HDFX28 closer to its rest--frame optical,
or if it were located at the modest redshifts of the catalogs described
above.

The opposite is true of $A$.  Star formation in the disk of a spiral is UV
bright, producing large measures of asymmetry in the UV even though the
galaxy may appear symmetric in the optical. \citet{kuchinski01} found that
$\Delta A$ between UV images and optical images of a galaxy can be as
large as $-0.7$ (again in the sense of $\Delta A = A_{\mbox{\tiny OPT}} -
A_{\mbox{\tiny FUV}}$), though $\Delta A \sim -0.3$ is more typical of
later spirals with moderate values of $A_{\mbox{\tiny FUV}}$
\citep[][figure 4]{kuchinski01}.

As for surface brightness dimming and the loss of spatial resolution,
\citet{kuchinski01} report that $C$ is in most cases robust to both
effects out to $z \sim 3$ ($\Delta C \simlt 5\%$), while the effect on $A$
is simply to increase its scatter ($\Delta A \simlt 12\%$). Consequently,
while we interpret bandshifting as resulting in a systematic shift in $C$
and $A$, we interpret the scatter introduced by surface brightness dimming
and the loss of spatial resolution as an increase in their error bars.
Hence, the morphological $k$--correction necessary to properly compare
HDFX28 to the artificially redshifted Frei catalog and to the galaxies of
the MDS amounts to an increase of $\sim 0.1$ in concentration index to
$\log C \sim -0.33$, with a corresponding increase in error bar to
$\sigma_{\mbox{\tiny $C$}} = 0.04$. Similarly, the asymmetry index must be
shifted down by $\sim -0.3$ to $\log A \sim -0.90$, with a corresponding
increase in error bar to $\sigma_{\mbox{\tiny $A$}} = 0.09$. Once applied,
these consideration show HDFX28 to fall definitively within the spiral
galaxies in the $\log C$ -- $\log A$ distribution of both samples
\citep{abraham96}. Therefore, we judge the morphology of HDFX28 to be
consistent with that of a rare, high--redshift spiral.

The inapplicability of the classical Hubble tuning fork to the galaxy
population at $z \gtrsim 0.5$ has been well--documented, and recent
morphological studies have shown the dearth of spirals at high redshift to
be a genuine change in the galaxy population --- not merely a function
cosmological distance effects on morphological classification
\citep[e.g.][]{vandenbergh02}. Possible scenarios posited to explain this
effect include the destruction of early--time disks from without by
mergers or from within by strong, starburst--driven galactic winds, or
perhaps stellar feedback in early disks suppresses the cooling of gas
before $z \sim 1$, preventing global dynamical instabilities from
initiating the formation of spiral structure \citep[][]{vandenbergh02rev}.
In any case, this single detection of an object at a redshift for which
spirals are not expected is certainly not a challenge to widely--accepted
hierarchical evolutionary scenarios, which have otherwise been successful
at predicting the results of deep galaxy surveys
\citep[e.g.][]{kauffmann93,baugh98}.  It may be the case that HDFX28 is
simply a rare example of an early--time disk which escaped destruction,
for instance, by a merger event.

\bigskip
\bigskip
\bigskip

\section{Conclusion}
\label{conclusion}

We have reported on two aspects of the high--redshift, hard X--ray
emitting spiral galaxy HDFX28: (1) its classification as a Type II AGN, a
population recently attracting renewed interest due to deep X--ray
surveys, and for which few {\it HST} images are available, and (2) its
unprecedented redshift for a galaxy with spiral morphology. As for HDFX28
as a Type II AGN, the canonical wisdom regarding weak, extended radio
sources with spectral indices steeper than $\alpha^{\mbox{\tiny 8.4
GHz}}_{\mbox{\tiny 1.4 GHz}} > 0.5$ dictates that such sources are driven
by star formation. Nonetheless, the combined weight of evidence from
X--ray, optical, and near--infrared observations of HDFX28 indicates the
presence of obscured AGN activity. It is instructive to note that when
re--interpreted in light of the spectroscopic redshift, even the
mid--infrared data for HDFX28 corroborates this result. At $z = 2.011$,
the ISOCAM LW3 filter samples rest wavelengths spanning only 4 $\mu$m to 5
$\mu$m. Here, the contribution to the mid--IR spectral energy distribution
made by UIB emission and by dust at 200 K is severely attenuated
\citep[see][Figure 1]{aussel99}.  Hence, the ISOCAM detection of this
source is far more plausibly explained by the hot, $\sim 10^3$ K dust
found in the central region of an AGN \citep[e.g.\ see][]{aussel98} than
it is by star formation alone.

As to the precise nature of the central engine in HDFX28, we conclude from
the comparatively narrow emission lines in the spectroscopy and from the
heavy obscuration evident in the X--ray data that HDFX28 is far more like
an obscured Type II system than an unobscured Type I system. Though this
conclusion is slightly at odds with the presence of weak, broad \hal\
emission, all remaining aspects of the source are entirely consonant with
observations of other Type II AGN at moderate--to--high redshifts
\citep[e.g.][]{kleinmann88, norman02, stern02q} and with HzRGs
\citep[e.g.][]{larkin00, mccarthy93, stern99hzrg, vernet01}.
\citet{norman02} describe a very similar situation in which their source
CDF--S 202 shows both the narrow ($\sim 1000$ km s$^{-1}$) emission lines
in its optical spectrum and the heavy obscuration in its X--ray emission
typical of a Type II system, but also shows emission line flux ratios
intermediate between Type I and Type II systems.  As noted by
\citet{stern02q}, it is conceivable that longer--wavelength spectra of
CDF--S 202 and other sources like it would also reveal broad \hal, though
they in every other way give evidence of the heavy obscuration considered
to be emblematic of Type II AGN.

Separately, the spectroscopy presented herein shows HDFX28 to be at an
unprecedented redshift for a galaxy with identifiably spiral structure.
Nevertheless, with the application of a modest morphological
$k$--correction, our quantitative analysis of its central concentration
and asymmetry is consistent with the interpretation that HDFX28 is a rare
example of a high--redshift spiral galaxy.  Owing to its proximity to the
HDF--N, HDFX28 will be subject to deep, space--based $B$, $V$, $i$, and
$z$ imaging with the Advanced Camera for Surveys as part of the upcoming
GOODS {\it HST} Treasury Program (M.\ Giavalisco, PI), as well as to
infrared imaging at $\lambda > 3$ $\mu$m with the Infrared Array Camera as
part of the GOODS {\it SIRTF} Legacy project (M.\ Dickinson, PI).  At a
minimum, the availability of multi--wavelength imaging will provide a
powerful additional lever arm on the issue of the morphology of HDFX28
\citep[e.g.][]{conselice97,conselice00}. As such, we eagerly look forward
to these expansive datasets.


\acknowledgments

We are grateful to Leonidas Moustakas, Mark Dickinson, Mauro Giavalisco, 
and the GOODS team for kindly providing the preliminary ACS imaging
of the HDF inner west flanking field.
In addition, we are indebted to the expert staff of the Keck Observatory, without whom
this work would not have been possible, and to J.~G. Cohen and C.~C.
Steidel for supporting LRIS--R and LRIS--B, respectively. We gratefully
acknowledge the careful reading and useful commentary of the anonymous
referee, by which this work substantially benefited. Finally, the authors
wish to acknowledge the significant cultural role that the summit of Mauna
Kea plays within the indigenous Hawaiian community. We are fortunate to
have the opportunity to conduct observations from this mountain. The work
of SD was supported by IGPP--LLNL University Collaborative Research
Program grant \#02--AP--015, and was performed under the auspices of the
U.S.  Department of Energy, National Nuclear Security Administration by
the University of California, Lawrence Livermore National Laboratory under
contract No.\ W--7405--Eng--48. The work of DS was carried out at the Jet
Propulsion Laboratory, California Institute of Technology, under contract
with NASA. HS gratefully acknowledges NSF grant AST 95--28536 for
supporting much of the research presented herein.  ML is grateful for
research support from the Beatrice Watson Parrent Fellowship at the
University of Hawai`i. This work made use of NASA's Astrophysics Data
System Abstract Service.




\begin{deluxetable}{ccccccl}
\tabletypesize{\scriptsize}
\tablewidth{0pt}
\tablecolumns{7}
\tablecaption{Emission--Line Measurements of HDFX28}
\tablehead{\colhead{Line} & \colhead{$\lambda_{\mbox{\tiny obs}}$} & \colhead{Redshift} & \colhead{Flux} &
\colhead{FWHM\tablenotemark{\dag}}
& \colhead{\wrst} & \colhead{Comment} \\ \colhead{ } & \colhead{(\AA)} & \colhead{ } &
\colhead{($10^{-17}$ ergs cm$^{-2}$ s$^{-1}$)} & \colhead{(km s$^{-1}$)} & \colhead{(\AA)} & \colhead{ }}
\startdata
\lya   & $3664.4  \pm 0.7$ & $2.0135 \pm 0.0006$ & $1.68 \pm 0.08$ & $1270 \pm 30$  & $35 \pm 3$  & LRIS--B \\
\nvw   & $3733.5  \pm 0.7$ & $2.0109 \pm 0.0006$ & $2.3  \pm 0.1$  & $2110 \pm 30$  & $50 \pm 3$  & LRIS--B \\
\civw  & $4665.1  \pm 0.7$ & $2.0117 \pm 0.0005$ & $2.47 \pm 0.08$ & $1300 \pm 10$  & $60 \pm 3$  & LRIS--B \\
\heiiw & $4935.8  \pm 0.7$ & $2.0096 \pm 0.0004$ & $0.45 \pm 0.08$ & $1400 \pm 60$  & $13 \pm 3$  & LRIS--B \\
\ciiiw & $5745.9  \pm 0.8$ & $2.0099 \pm 0.0004$ & $0.19 \pm 0.08$ & $900  \pm 130$ & $7  \pm 3$  & LRIS--B \\
\neivw & $7292.9  \pm 0.7$ & $2.086  \pm 0.0003$ & $0.3  \pm 0.1$  & $1470 \pm 30$  & $9  \pm 7$  & LRIS--R \\
\niiaw & $19835.2 \pm 0.7$ & $2.0155 \pm 0.0001$ & $0.24 \pm 0.08$ & $380  \pm 30$  & $9  \pm 11$ & NIRSPEC \\
\haln  & $19790.5 \pm 0.7$ & $2.0155 \pm 0.0001$ & $0.7  \pm 0.1$  & $240  \pm 30$  & $26 \pm 9$  & NIRSPEC \\
\halb  & $19780   \pm 8$   & $2.014  \pm 0.001$  & $2.3  \pm 0.3$  & $2500 \pm 250$ & $90 \pm 30$ & NIRSPEC \\
\niibw & $19897.3 \pm 0.7$ & $2.0155 \pm 0.0001$ & $0.72 \pm 0.09$ & $380  \pm 30$  & $30 \pm 12$ & NIRSPEC \\
\enddata
\tablenotetext{\dag}{The line widths have been deconvolved according to
the relation $\mbox{FWHM(obs)}^2 = \mbox{FWHM(instr)}^2 + \mbox{FWHM(inher)}^2$,
where FWHM(obs) is the observed line width, FWHM(instr) is the instrumental
resolution, and FWHM(inher) is the inherent line width.}
\tablecomments{The uncertainties quoted in this table are dominated by
four sources of error: the statistical error due to Poisson noise in the
spectrum, the readnoise due to the detector, a systematic error
introduced by sky subtraction during the data processing, and the $1
\sigma$ uncertainties derived for the fit parameters.  All of these errors
are easily characterized except for the systematic error introduced by sky
subtraction. Consequently, we assumed that this additional error is at
least as large as the statistical error, and we added it in quadrature to
form the total uncertainty.
}
\label{lines}
\end{deluxetable}


\begin{deluxetable}{lccc}
\tablewidth{0pt}
\tablecolumns{2}
\tablecaption{Diagnostic Emission--Line Ratios}
\tablehead{\colhead{Diagnostic} & \multicolumn{3}{c}{Flux Ratio} \\
& \colhead{HDFX28} & \colhead{CDF--S 202\tablenotemark{1}} & \colhead{CXO52\tablenotemark{2}}}
\startdata
\lya\ / \civw\  & $0.7 \pm 0.1$ & 1.66    & $5.4 \pm 0.4$ \\
\lya\ / \haln\  & $2.4 \pm 0.1$ & \nodata & \nodata       \\
\nvw\ / \lya    & $1.4 \pm 0.1$ & 0.36    & 0.03:         \\
\nvw\ / \civw   & $0.9 \pm 0.1$ & 0.60    & 0.2:          \\
\nvw\ / \heiiw  & $5.2 \pm 0.1$ & 2.11    & 0.4:          \\
\civw\ / \heiiw & $5.5 \pm 0.1$ & 3.54    & $2.1 \pm 0.3$ \\
\niibw\ / \haln & $1.0 \pm 0.1$ & \nodata & \nodata       \\
\enddata
\tablenotetext{1}{Source: \citet{norman02}.}
\tablenotetext{2}{Source: \citet{stern02q}.}
\label{ratios}
\end{deluxetable}


\begin{deluxetable}{cccclc}
\tabletypesize{\scriptsize}
\tablewidth{0pt}
\tablecolumns{6}
\tablecaption{Photometry of HDFX28}
\tablehead{\colhead{Observed} & \colhead{Rest frame} & \colhead{Observed} & \colhead{Flux Density} &
\colhead{Detector/} & \colhead{Reference} \\ \colhead{Bandpass} & \colhead{Central $\lambda$}
& \colhead{Magnitude\tablenotemark{\dag}} &
\colhead{($\mu$Jy)} & \colhead{Instrument}}
\startdata
2--8 keV          & 18.1 keV   & \nodata & $12 \times 10^{-5}$\tablenotemark{\ddag}
& {\em Chandra}/ACIS        & 1 \\
0.5--2 keV        & 3.8 keV    & \nodata & $5 \times 10^{-5}$\tablenotemark{\ddag}
& {\em Chandra}/ACIS        & 1 \\
Harris $U$        & 1200 \AA   & 24.40   & 0.33                  & KPNO 4m/MOSAIC            & 2 \\
$B$               & 1450 \AA   & 24.05   & 1.02                  & Keck/LRIS                 & 3 \\
$G$               & 1600 \AA   & 24.37   & 0.69                  & Palomar 200--inch/COSMIC  & 4 \\
Kron--Cousins $V$ & 1830 \AA   & 23.89   & 1.01                  & CFHT/Hawaii 8K CCD Mosaic & 5 \\
Kron--Cousins $R$ & 2160 \AA   & 23.5    & 1.2                   & Keck/LRIS                 & 3 \\
$\cal{R}$         & 2300 \AA   & 23.75   & 0.87                  & Palomar 200--inch/COSMIC  & 4 \\
Kron--Cousins $I$ & 2660 \AA   & 22.9    & 1.8                   & CFHT/Hawaii 8K CCD Mosaic & 3 \\
$I_{814}$(AB)     & 2700 \AA   & 23.46   & 1.43                  & {\em HST}/WFPC2           & 6 \\
$HK^\prime$       & 5980 \AA   & 19.3    & 9.7                   & UH 2.2m/QUIRC             & 5 \\
$K_s$             & 7140 \AA   & 19.01   & 17.62                 & Palomar 200--inch/COSMIC  & 4 \\
15 $\mu$m         & 5.0 $\mu$m & \nodata & 441$^{+43}_{-82}$     & {\em ISO}/ISOCAM          & 7 \\
8.5 GHz           & 25.6 GHz   & \nodata & 8.15                  & VLA                       & 8 \\
1.4 GHz           & 4.2 GHz    & \nodata & 87.8                  & VLA                       & 9 \\
\enddata
\tablenotetext{\dag}{All magnitudes are normalized to Vega, except the $I_{814}$ magnitude,
which is AB.  The conversion between Vega--based $I$--band magnitudes and $I_{\mbox{\tiny AB}}$
is $I \approx I_{\mbox{\tiny AB}} - 0.3$.}
\tablenotetext{\ddag}{X--ray flux densities were estimated from the {\em Chandra}/ACIS hard
and soft band fluxes by assuming an X--ray spectral index of $\alpha = +0.7$
(where $F(E) \propto E^\alpha$), based on the hardness ratio described in \S~\ref{photometry}.}
\tablerefs{(1) \citealt{hornschemeier01};
(2) C. McNally et al., in preparation;
(3) \citealt{barger00};
(4) \citealt{hogg00};
(5) \citealt{barger99};
(6) \citealt{mann97};
(7) \citealt{aussel99};
(8) \citealt{richards98};
(9) \citealt{richards00}.}
\label{photo}
\end{deluxetable}

\end{document}